\documentclass[conference]{IEEEtran}
\usepackage{stackengine}

\IEEEoverridecommandlockouts
\usepackage{cite}
\usepackage{amsmath,amssymb,amsfonts}
\usepackage{graphicx}
\usepackage{xcolor}
\def\BibTeX{{\rm B\kern-.05em{\sc i\kern-.025em b}\kern-.08em
		T\kern-.1667em\lower.7ex\hbox{E}\kern-.125emX}}
\usepackage{epsfig}
\usepackage{theorem}
\usepackage{mathrsfs}
\usepackage{diagbox}
\usepackage[utf8]{inputenc}
\usepackage[T1]{fontenc}
\usepackage{newtxtext,newtxmath}
\usepackage{pdfrender}
\usepackage{subfigure}
\usepackage{caption}

\DeclareMathOperator*{\argmin}{arg\,min}
\usepackage{csquotes}
\newcommand*{\boldgreek}[1]{%
	\textpdfrender{%
		TextRenderingMode=FillStroke,%
		LineWidth=.35pt,%
	}{#1}%
}

\usepackage{mathrsfs}
\usepackage{euscript}
\usepackage{graphicx}
\usepackage{multirow}
\usepackage{algorithmicx}
\usepackage{algpseudocode}

\usepackage{cite}
\usepackage{url}
\usepackage[export]{adjustbox}
\usepackage{algorithm}
\usepackage[letterpaper, top=0.7in, bottom=1.02in, left=0.673in, right=0.673in, twocolumn, columnsep=0.161in]{geometry}

\begin{document}

\title{
Meta Reinforcement Learning for Resource Allocation in Multi-Antenna UAV Network with Rate Splitting Multiple Access
\vspace{-0.50cm}}
\author{\IEEEauthorblockN 
{Hosein~Zarini$^{\dag}$, Maryam~Farajzadeh~Dehkordi$^{\S}$, Armin~Farhadi$^{\S\S,\dag\dag}$,~Mohammad\! Robat  \!Mili$^{\star}$,\\~Ali~Movaghar$^{\dag}$,~Mehdi~Rasti$^{\ddagger\ddagger}$,~Yonghui~Li$^{\ddagger}$~and~Kai-Kit~Wong$^{\dag\dag}$
}\\$^{\dag}$Dept. of Computer Engineering, Sharif University of Technology, Tehran, Iran
		\\$^{\S}$Dept. of Electrical and Computer Engineering, George Mason University, VA, USA\\
		$^{\S\S}$Dept. of Electrical and Computer Engineering, University of Tehran, Tehran, Iran\\
        $^\star$Pasargad Institute for Advanced Innovative Solutions (PIAIS), Tehran, Iran\\       $^{\ddagger\ddagger}$Centre for Wireless Communications, University of Oulu, Oulu, Finland\\$^{\ddagger}$School of Electrical and Information Engineering, The University of Sydney, Sydney, Australia\\
        $^{\dag\dag}$Dept. of Electronic and Electrical Engineering, University College London, London, United Kingdom
        \vspace{-0.55cm}\\
	}
\maketitle\vspace{-0.20cm}
\begin{abstract} 
Unmanned aerial vehicles (UAVs) with multiple antennas have recently been explored to improve capacity in wireless networks. However, the strict energy constraint of UAVs, given their simultaneous flying and communication tasks, renders the exploration of energy-efficient multi-antenna techniques indispensable for UAVs. Meanwhile, lens antenna subarray (LAS) emerges as a promising energy-efficient solution that has not been previously harnessed for this purpose. In this paper, we propose a LAS-aided multi-antenna UAV to serve ground users in the downlink transmission of the terahertz (THz) band, utilizing rate splitting multiple access (RSMA) for effective beam division multiplexing. We formulate an optimization problem of maximizing the total system spectral efficiency (SE). This involves optimizing the UAV's transmit beamforming and the common rate of RSMA. 
By recasting the optimization problem into a Markov decision process (MDP), we propose a deep deterministic policy gradient (DDPG)-based resource allocation mechanism tailored to capture problem dynamics and optimize its variables. Moreover, given the UAV's frequent mobility and consequential system reconfigurations, we fortify the trained DDPG model with a meta-learning strategy, enhancing its adaptability to system variations. Numerically, more than 20\% energy efficiency gain is achieved by our proposed LAS-aided multi-antenna UAV equipped with 4 lenses, compared to a single-lens UAV. Simulations also demonstrate that at a signal-to-noise (SNR) of 10 dB, the incorporation of RSMA results in a 22\% SE enhancement over conventional orthogonal beam division multiple access. Furthermore, the overall system SE improves by 27\%, when meta-learning is employed for fine-tuning the conventional DDPG method in literature.
%Unmanned aerial vehicles (UAVs) with multiple antennas have recently been investigated for capacity improvement in wireless networks. Nonetheless, strict energy limitations of UAVs for flying and communication at the same time makes the investigation of green multi-antenna techniques for UAVs, indispensable. Lens antenna subarray (LAS) is such a potential energy efficient candidate, yet never tried to this purpose before. In this paper, we propose a LAS-aided multi-antenna UAV for serving ground users in downlink transmission of terahertz band and employ rate splitting multiple access (RSMA) for effective beam division multiplexing. To do so, an optimization problem is formulated aimed at maximizing the total system spectral efficiency (SE), for optimizing the transmit beamforming of the UAV and the common rate of RSMA, which respects the quality-of-service (QoS) for ground users and the power budget of the UAV. 
%Incorporating coupled variables makes this problem extremely sophisticated, to deal with. Due to mobility of the UAV, a dynamic radio resource allocation is proposed by a fine-tuned deep deterministic policy gradient (DDPG) method, whose parameters are initialized via meta learning for capturing the dynamicity of system and faster convergence. Simulations reveal that for SNR=10dB, incorporating RSMA achieves 22\% SE enhancement over the conventional orthogonal beam division multiple access. Additionally, the overall system SE is improved by 27\%, when meta learning is used for fine-tuning the DDPG.
\end{abstract}
\begin{IEEEkeywords}
Unmanned aerial vehicle (UAV), lens antenna subarray (LAS), rate splitting multiple access (RSMA), beam division multiplexing, deep deterministic policy gradient (DDPG), meta learning\vspace{-0.20cm}
\end{IEEEkeywords}
\IEEEpeerreviewmaketitle
\vspace{-0.15cm}
\section{Introduction}
\vspace{-0.15cm}
\subsection{Background and Incentives}
\par Unmanned aerial vehicles (UAVs), known for their flexible maneuverability, readiness for use, lightweight, low cost, and simple handling stand out as a pioneering technology envisioned for the upcoming generation of wireless networks. In this context, massive multiple-input multiple-output (MIMO) techniques, as potential capacity enablers, are advocated for deployment on UAVs\cite{3D}. 
The studies in the realm of multi-antenna UAVs are primarily on the consideration of energy consumption. Specifically, the substantial energy required in the conventional fully digital MIMO renders it impractical for UAV deployment. Consequently, energy-efficient multi-antenna alternatives like beamspace MIMO have been proposed\cite{H2,H3,H4}, holding the potential for delivering seamless and sustainable aerial communication. 
\par Recently, the concept of a single-lens beamspace MIMO was explored for UAVs~\cite{BU}. Despite its notable energy savings, this single-lens architecture grapples with several deficiencies, including power leakage issues, a large focal depth, and the lens size relative to the number of antennas. Addressing these drawbacks, the work in \cite{LAS1} introduced a more efficient MIMO architecture, termed lens antenna subarray (LAS), which replaces the large single-lens with multiple smaller ones. This design aims to curtail power consumption and hardware complexity, wherein each radio frequency (RF) chain is connected to a lens via a single phase shifter (PS) and switch, enhancing energy efficiency significantly\cite{LAS2}. With these advancements, integrating this technology into UAVs appears promising for green and sustainable aerial communication. Meanwhile, although the resource allocation aspect of a LAS system was investigated in \cite{LAS3}, a review of previous studies \cite{LAS1,LAS2,LAS3} indicates that they all share a common limitation: they adhere to strict resource orthogonality, severely limiting the number of users these systems can support. Herein lies the potential of rate splitting multiple access (RSMA), capable of transcending the fundamental constraints of resource orthogonality and facilitating massive connectivity\cite{RSMA}. This approach could notably augment the performance of orthogonal LAS systems\cite{LAS1,LAS2,LAS3}.
\vspace{-0.15cm}
\subsection{Research Challenges and Contributions}
Reviewing literature unveils that on one hand, the use of multi-antenna techniques considerably enhances the data rate of UAV-enabled networks \cite{MIMO1,MIMO2,MIMO3}. The energy consumption corresponded to a multi-antenna device, on the other hand, scales up in orders of magnitude, compared to that for a single antenna device. 
\par Regarding these challenges, we investigate the use of LAS as a green multi-antenna structure for UAV in this paper. In comparison with\cite{BU}, where the UAV is equipped with a single-lens beamspace multi-antenna system, a LAS-enabled UAV employs a set of smaller lenses that consumes less energy. Additionally, compared to existing LAS systems \cite{LAS1,LAS2,LAS3}, we employ RSMA for downlink beam division multiplexing that brings about more efficient interference control and thereby data rate improvement\cite{RSMA}. Toward optimizing multi-antenna UAV networks, various resource allocation frameworks have been proposed to date \cite{MIMO1,MIMO2,MIMO3}, none of them however render well-suited in our system. While classical iterative approaches \cite{MIMO1,MIMO2} are computationally intensive, the class of well-investigated deep reinforcement learning (DRL) methods \cite{MIMO3} already fail in continuously-changing wireless systems, wherein the rate of environment variations dominates the interval between the training and testing stages of the DRL. Concerning the high mobility of the UAV and thereby our frequently varying wireless system, we propose a real-time resource allocation scheme relying on model-agnostic meta learning~\cite{meta}. The proposed solution captures the environment dynamicity and exhibits significant generalization ability to adapt to upcoming situations.
Simulations indicate energy and spectral efficiency superiority over a a single-lens beamspace UAV \cite{BU}.
\par The remainder of this paper is organized as follows. Section \ref{SM} describes the system setup, signal model and problem statement. Section \ref{SS} introduces the solution strategy and its complexity analysis. Eventually, simulation results and conclusions are discussed in Sections \ref{SR} and \ref{conclutions}, respectively.
\vspace{-0.25cm}
\section{System Setup}\label{SM}
\vspace{-0.0cm}
\subsection{LAS Architecture}
\begin{figure}
%\centering
\hspace{-4.7cm}\includegraphics[width=17.0cm,height=6.0cm]{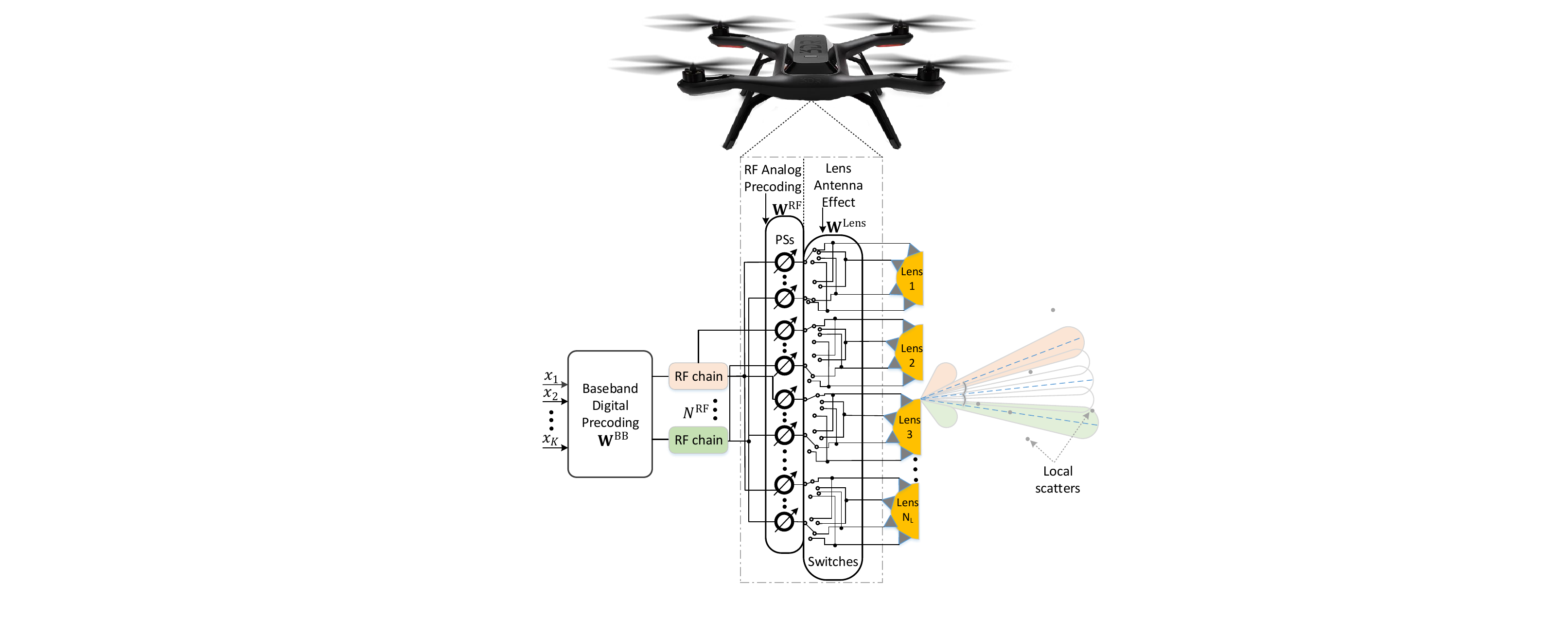}
\caption{A UAV with LAS.}
\label{frame}
\vspace{-0.60cm}
\end{figure}

\vspace{-0.05cm}
As shown in Fig.~1, consider an UAV with coordinates of $\mathbf{{q}}_{U}=(x_U,y_U)$, hovering at the altitude of $z_U$ and employing a LAS system with $N_{\textrm{t}}$ antennas and $N_{\textrm{Lens}}$ lenses, each equipped with $M={N_{\textrm{t}}/N_{\textrm{Lens}}}$ antenna elements to serve $K$
single-antenna users. Compared to the distance between the UAV and ground users, the distance between the lenses at the UAV is negligible. Therefore, a unified coordinate is considered for the multi-lens UAV. The UAV is also assumed to be equipped with $N^{\textrm{RF}}$ RF chains and also a uniform planar array (UPA) with $N_{\textrm{x}}$ and $N_{\textrm{y}}$ antenna elements in horizontal and vertical directions, respectively, such that $N_{\textrm{t}}=N_{\textrm{x}}N_{\textrm{y}}$ and $N^{\textrm{RF}}$\(\ll\)$N_{\textrm{t}}$. According to Fig.~\ref{frame}, the user information symbols first of all get precoded through a baseband digital precoder matrix, denoted by $\textbf{W}^\textrm{\textbf{BB}}\in \mathbb{C}^{N^{\textrm{RF}}\times K}$. By doing so, the signals reside in the RF domain using RF chains. As each RF chain is connected into a specific lens, further precoding is performed through a $\textbf{W}^\textrm{{RF}}\in \mathbb{C}^{N_{\textrm{Lens}}\times N^{\textrm{RF}}}$ RF analog precoder with low-resolution $B$-bit PSs. Finally, the analog domain signals are routed through a switching system,
characterized by an antenna affect precoder $\textbf{W}^\textrm{{Lens}}\in \mathbb{C}^{N_{\textrm{t}}\times N_{\textrm{Lens}}}$, which is stated as:
\begin{align}
\mathbf{W}^{\text{Lens}}=\left[ 
\begin{array}{cccc}
{\boldgreek{\Gamma}}_{1} & \mathbf{0}_{{M}\times 1} & \cdots  & 
\mathbf{0}_{{M}\times 1} \\ 
\mathbf{0}_{{M}\times 1} & {\boldgreek{\Gamma}}_{2} &  & \mathbf{0}%
_{{M}\times 1} \\ 
\vdots  &  & \ddots  & \vdots  \\ 
\mathbf{0}_{{M}\times 1} & \mathbf{0}_{{M}\times 1} &  & {\boldgreek{\Gamma}}_{N_{\textrm{Lens}}}%
\end{array}%
\right]_{N_{\textrm{t}}\times N_{\textrm{Lens}}},
\end{align}
where $\Gamma_l\in \mathbb{C}^{M\times 1}$ specifies the indices of selected beams at the $l$th lens. 
%The direct current (DC) power consumption of a LAS-equipped transmitter before transmission of wireless signals would be modelled as\cite{LAS2}: $P^{\textrm{LAS}}=P^{\textrm{Circ}}+\dfrac{P^{\textrm{RF}}}{\eta_{\textrm{PA}}\eta_\textrm{SW}}$, in which $P^{\textrm{RF}}$ denotes the total RF power to be transmitted by the antenna elements; $\eta_{\textrm{PA}}$ and $\eta_\textrm{SW}$ respectively stand for the power amplifier efficiency and the efficiency of the switches and $P^{\textrm{Circ}}$ specifies the circuit power. In particular, $\eta_\textrm{SW}=10^{-{\zeta\textrm{IL}_{\textrm{SW}}}/10}$, with $\zeta$ and $\textrm{IL}_{\textrm{SW}}$, respectively being the number of required series switches under each lens and the switch insertion loss. Besides, the circuit power can be modelled as $P^{\textrm{Circ}}=\textrm{N}^{\textrm{RF}}\textrm{N}_{\textrm{Lens}}P^{\textrm{PS}} +\zeta\textrm{N}^{\textrm{RF}}\textrm{N}_{\textrm{Lens}}P^{\textrm{SW}}+\textrm{N}^{\textrm{RF}}P^{\textrm{RF}}$ where $P^{\textrm{PS}}$, $P^{\textrm{SW}}$ and $P^{\textrm{RF}}$ represent the power consumption per PS, switch and RF chain, respectively.
\subsection{Signal Model}
Let us denote by $S_{j}~\forall j\in\{1,2,$\textperiodcentered
\textperiodcentered \textperiodcentered $,J\}$, the set of users served
by the $j$th beam, where $S_i\cap S_j=\Phi, $~ $i\neq j~\forall i,j\in\{1,2,$\textperiodcentered
\textperiodcentered \textperiodcentered $,J\}$, $%
\sum_{j=1}^{J}|S_{j}|$ $=K$ and $J$ indicates the total number of beams. It is supposed that user clustering is carried out in a predefined manner at each beam. More precisely, each user is served via a specific beam at most, and user-beam assignments are performed on the basis of channel conditions. That is to say, each beam $j$ covers $S_j$ users with better channel conditions from its perspective.
%The messages $(x_{1},x_{2},...,x_{K})$  intended to be sent to $K$ users of two parts in RSMA, including the common messages $%(x_{c,1},x_{c,2},...,x_{c,K})$ as well as the private messages $(x_{p,1},x_{p,2},...,x_{c,K})$.
%The common parts of all users are combined into one message, denoted by $X_{(\textrm{C})}$ and then encoded into a a unified common stream denoted by $s^{(\textrm{C})}$, with a well-known codebook to all users. The private parts on the other hand, are encoded separately into private streams $(s_{1},s_{2},...,s_{K})$ and superimposed over the common message using linearly precoding at each beam. 
The UAV is presumed to leverage RSMA, through which each user message is split into common and private parts before the transmission. Whereas the common message has the same content for all users, each user has its specific private message with unique content. By integrating the user messages for $K$ users into a superimposed signal, that will carry $K+1$ messages: one common message for all users, in addition to $K$ private messages, each corresponds to one user. In light of the above clarification, the superimposed precoded transmit signal will be 
$
\mathbf{s}=\sqrt{P^{(\textrm{C})}}\textbf{W}^{\textrm{{Lens}}}\textbf{W}^{\textrm{{RF}}}\textbf{w}^{\textrm{{BB},(\textrm{C})}}s^{(\textrm{C})}+\sum_{j=1}^{N_{\textrm{t}}}\sum_{k=1}^{|S_{j}|}\sqrt{P_{j,k}}\textbf{W}^\textrm{{Lens}}\textbf{W}^\textrm{{RF}}\textbf{w}^\textrm{{BB}}_{k}s_{j,k},
$
with $P^{(\textrm{C})}$ expressing the transmit power for the common stream, $P_{j,k}$ denoting the transmit power for the private stream of the $k$th user from the $j$th beam; $s^{(\textrm{C})}$ standing for the transmit symbol for the common stream and finally $s_{j,k}$ indicating the transmit symbol from the $j$th beam for the ground user $k$ with normalized power $\mathbb{E}{\{|s_{j,k}|^{2}}\}=1$. 
%The common stream $s^{(\textrm{C})}$ at each user in the $j$th beam is decoded first into $\hat{X}_{(\textrm{C})}$, for which the interference from the private streams is regarded as noise. Relying on SIC, each user in the $j$th beam will re-encode $\hat{X}_{(\textrm{C})}$, precode it then, and eliminate it from its received signal. Under the assumption of residual interference from the remaining private stream as noise, each user $k$ will then decode its private stream from the $j$th beam, i.e., $s_{j,k}$ into $\hat{x}_{p,k}$. Ultimately, the $k$th user in the $j$th beam restores its original message by eliminating $\hat{x}_{c,k}$ from $\hat{X}_{(\textrm{C})}$, and merging $\hat{x}_{c,k}$ with $\hat{x}_{p,k}$, resulting in $\hat{x}_{k}$. As shown in Fig. , this process is defined as 1-layer rate splitting thanks to incorporating a single common message, as well as a single SIC layer at each user. 
\par Accordingly, the narrow-band received signal at the $k$th ground user from the $j$th beam in such a system can be expressed as
$
{y}_{j,k}=~{h}^{H}_{j,k}\sqrt{P^{(\textrm{C})}}\textbf{W}^\textrm{{Lens}}\textbf{W}^\textrm{{RF}}\textbf{w}^{\textrm{{BB},(\textrm{C})}}s^{(\textrm{C})}\!+{h}^{H}_{j,k}\sum_{j=1}^{N_{\textrm{t}}}\sum_{k=1}^{|S_{j}|}%
\sqrt{P_{j,k}}\textbf{W}^\textrm{{Lens}}\textbf{W}^\textrm{{RF}}\textbf{w}^\textrm{{BB}}_{k}s_{j,k}+{\nu}_k,$
where ${\nu}\sim\mathcal{CN}(0,\sigma_{\nu}^{2})$ indicates the independent and identically distributed (i.i.d.) complex Gaussian
noise, in which $\sigma_{\nu}$ denotes the noise power. Furthermore, $h_{j,k}$ introduces the channel response from the $j$th beam to the $k$th ground user with a half-wavelength-spacing UPA. By leveraging existing efficient channel estimation techniques\cite{H4}, the channel matrix between the UAV and users can be efficiently obtained, which is denoted by $\mathbf{H}\in{\mathbb{C}^{K\times N_{\textrm{t}}}}$, where $\mathbf{H}=[\mathbf{h}_{1},\mathbf{h}_{2},...,\mathbf{h}_{K}]$. In particular, we consider the terahertz (THz) band for transmission and $\mathbf{h}_{k}\in{\mathbb{C}^{1\times N_{\textrm{t}}}}$ introduces the channel response vector for the $k$th ground user, defined as:
$
\mathbf{h}_{k}=\sum_{l=0}^{L_k}\beta _{k}^{(l)}(\tilde{d}_k)\Dot{\mathbf{a}}\left( \theta _{k}^{(l)},\psi _{k}^{(l)}\right),
$
wherein $l=0$ corresponds to the line-of-sight (LoS) path, while all $l>0$ correspond to the non-line-of-sight (NLoS) ones; $\beta _{k}^{(l)}(\tilde{d}_k)$ is the distance-based complex-valued gain %\cite{Zhang}  
with $\tilde{d}_k$ indicating the distance between the UAV and the ground user $k$; As well, $\theta_{k}^{(l)}$ and $\psi_{k}^{(0)}$, respectively stand for the azimuth and elevation angle of departure (AoD).
Besides, the $1\times N_{\textrm{t}}$ array steering vector $\Dot{\mathbf{a}}\left( \theta _{k}^{(l)},\psi _{k}^{(l)}\right)$ can be declared as: 
$
\Dot{\mathbf{a}}\left( \theta,\psi \right) =\frac{1}{\sqrt{N_{\textrm{t}}}}\left[ e^{-j2\pi d \sin(\theta)\sin(\psi)n_1/\lambda}%
\right]\otimes\left[ e^{-j2\pi d cos(\psi)n_2/\lambda}%
\right],
$
in which $j=\sqrt{-1}$; $n_1=\left[0,1,\text{\textperiodcentered }\ \text{%
\textperiodcentered }\ \text{\textperiodcentered }\ ,N_x-1\right] $, $n_2=\left[0,1,\text{\textperiodcentered }\ \text{%
\textperiodcentered }\ \text{\textperiodcentered }\ ,N_y-1\right] $, $\lambda$ denotes the signal carrier wavelength and $d = \lambda/2$ indicates the antenna spacing. Compared to the amplitude of the LoS component $|\beta _{k}^{(0)}(\tilde{d}_k)|$, the amplitude of NLoS components $ |\beta _{k}^{(l)}(\tilde{d}_k)|~\forall l\ge{1}$, at THz frequencies are much weaker and thus, the LoS link is solely taken into consideration. Each ground user $k$ is supposed to be of the coordinates $\mathbf{{q}}_{k}=(x_k,y_k)$ with zero altitude. Then, the ground coverage
area of the UAV depends on the antenna’s main lobe. Therefore, $||\mathbf{{q}}_{U}-\mathbf{q}_{k}||_{2}=z_U\tan\psi^{(0)}_{k}$ and $y_U-y_k=(x_U-x_k)\tan\theta^{(0)}_{k}$\cite{3D}. On this basis, $\tilde{d}_k=\sqrt{||\mathbf{{q}}_{U}-\mathbf{q}_{k}||_{2}^{2}+z_{U}^{2}}$\cite{BU}.
\par Regarding RSMA, each user at the first stage decodes the common stream by considering the private streams, all as noise. Next, the private stream for each user can be obtained by applying successive interference cancellation (SIC), through which the common stream will be removed from its observation, whereas other private streams are treated as noise.
%By indicating the coordinates of the UAV as $\mathbf{{q}}_{U}=(x,y,z)$, the azimuth and elevation AoDs for the LoS path can be accordingly restated as $\psi^{(0)}=\arctan \dfrac{\sqrt{x^{2}+y^{2}}}{z}$ and $\theta^{(0)}=\arctan \dfrac{y}{x}$\cite{3D}.
\subsection{Problem Statement}
In a LAS-assisted system, multi-level precoding plays an important role in overall system performance\cite{LAS2}. Thus, it is of primary importance to meticulously optimize the transmit precoding in such a system. First consider the digital baseband precoding $\textbf{W}^\textrm{{BB}}$. 
Once the channel state information (CSI) is estimated, zero-forcing method can be effectively applied to the baseband digital precoding so as to suppress the inter-beam interference, defined as:
$
    \textbf{W}^\textrm{{BB}} = \Bigg({\textbf{W}^\textrm{{RF}}}^{H}{\textbf{W}^\textrm{{Lens}}}^{H}\textbf{H}^{H}\textbf{H}\textbf{W}^\textrm{{Lens}}\textbf{W}^\textrm{{RF}}\Bigg)^{-1}\Bigg(\textbf{H}\textbf{W}^\textrm{{Lens}}\textbf{W}^\textrm{{RF}}\Bigg)^{H}.
$
Next, for the RF analog precoder, all the PSs are commonly supposed to have the same amplitude of $\dfrac{1}{\sqrt{N_{\textrm{Lens}}}}$\cite{LAS2}, yet equipped with different phases. Hence, this precoder for a typical RF $f$ with $B$-bit PSs, can be characterized as follows\cite{LAS3}:
$
\textbf{w}^\textrm{{RF}}_{f}=\dfrac{1}{\sqrt{N_{\textrm{Lens}}}}\bigg\{e^{j\frac{2\pi b}{2^B}}\big|b=0,1,...,2^{B}-1\bigg\}.$
Taking RSMA into account again, the common received data rate of the $k$th user associated with the $j$th beam based on what discussed hitherto, can be expressed as: $
    R^{(\textrm{C})}_{j,k}({P}^{(\textrm{C})},\mathbf{{q}}_{U},\textbf{W}^\textrm{{Lens}})=\log_{2}\bigg(1+\dfrac{|\mathbf{h}_{j,k}^{H}\textbf{W}^\textrm{{Lens}}\textbf{W}^\textrm{{RF}}\textbf{w}^\textrm{{BB}}_{k}|_{2}^{2}P^{(\textrm{C})}}{I_{j}+\sigma^{2}_{k}}\bigg),$
in which $P^{(\textrm{C})}$ denotes the transmit power of the UAV for the common stream, and $
    I_{j}=\sum_{k=1}^{|S_{j}|}|h^{H}_{j,k}\textbf{W}^\textrm{{Lens}}\textbf{W}^\textrm{{RF}}\textbf{w}^\textrm{{BB}}_{k}|^{2}P_{j,k}.$
On the other hand, the private received data rate of the $k$th user associated with the $j$th beam can be expressed as $
    R^{(\textrm{P})}_{j,k}(\textbf{P}^{(\textrm{P})},\mathbf{{q}}_{U},\textbf{W}^\textrm{{Lens}})=\log_{2}\bigg(1+\dfrac{|\mathbf{h}_{j,k}^{H}\textbf{W}^\textrm{{Lens}}\textbf{W}^\textrm{{RF}}\textbf{w}^\textrm{{BB}}_{k}|_{2}^{2}P_{j,k}}{I^{\prime}_{j}+\sigma^{2}_{k}}\bigg),
$
where $\textbf{P}^{(\textrm{P})}=[P_{j,k}]_{K\times J}$ and
$
    I^{\prime}_{j}=\sum_{k^{\prime}\neq k, k^{\prime}=1}^{|S_{j}|}|h^{H}_{j,k}\textbf{W}^\textrm{{Lens}}\textbf{W}^\textrm{{RF}}\textbf{w}^\textrm{{BB}}_{k}|^{2}P_{j,k^{\prime}}.
$
In order to decode the common stream at all users successfully, it is necessary to guarantee a common receiving data rate for each user $k$ from the $j$th beam, denoted by ${R}^{*}_{k}$, such that $\min_{k} {R^{(\textrm{C})}_{j,k}}({P}^{(\textrm{C})},\mathbf{{q}}_{U},\textbf{W}^\textrm{{Lens}})\ge \sum_{k=1}^{K} {R_{j,k}^{*}}.$
Now, the overall data rate of the user $k$ (including both common and private data rates) from the $j$th beam can be expressed as $\Tilde{R}_{j,k}(\textbf{P},\mathbf{{q}}_{U},\textbf{W}^\textrm{{Lens}},\mathbf{R}^{*})=R_{j,k}^{*}+{R^{(\textrm{P})}_{j,k}}(\textbf{P}^{(\textrm{P})},\mathbf{{q}}_{U},\textbf{W}^\textrm{{Lens}}),$
in which $\mathbf{{P}}=[P^{(\textrm{C})},P_{j,k}]_{J\times K+1}$ and $\textbf{R}^{*}=[R^{*}_{j,k}]_{J\times K}$.

\par 
%Owing to the energy limitation of the UAV as a flying serving BS, it is crucial to precisely optimize its transmit power as well as motion trajectory so as to provide a seamless and durable aerial communication. In this regard, 
Finally, based on what discussed so far, we formulate a network-wide radio resource allocation problem aimed at maximizing the total system spectral efficiency, which is posed as follows:
\vspace{-0.3cm}\begin{subequations}
\label{QoE_max_prob.1}
\begin{align}
\mathcal{P}_{1}:\nonumber&\max_{\{\textbf{P},\mathbf{{q}}_{U},\textbf{W}^\textrm{{Lens}},\mathbf{R}^{*}\}}\sum\limits_{j=1}^{J}\sum\limits_{k=1}^{\left\vert
S_{j}\right\vert }\Tilde{R}_{j,k}(\textbf{P},\mathbf{{q}}_{U},\textbf{W}^\textrm{{Lens}},\mathbf{R}^{*})
\\&\text{s.t.}~~~~P^{\textrm{LAS}}+P^{\textrm{Hov}}+\sum\limits_{j=1}^{J}\sum\limits_{k=1}^{\left\vert
S_{j}\right\vert }P_{j,k}\le {{P}^{\textrm{max}}},\\&~~~~~~~\Tilde{R}_{j,k}(\textbf{P},\mathbf{{q}}_{U},\textbf{W}^\textrm{{Lens}},\mathbf{R}^{*})\ge {\tilde{R}^{\textrm{min}}} ~~\forall j,k,
\\&~~~~~~~\mathbf{q}^{\textrm{min}}\le \mathbf{q}_U\le \mathbf{q}^{\textrm{max}},
\\&~~~~~~~\mathbf{q}_{U}(t+1)- \mathbf{q}_U(t)\le \mathrm{V}^{\textrm{max}},~~\forall t,
\\&~~~~~~~\min_{k} {R^{(\textrm{C})}_{j,k}}({P}^{(\textrm{C})},\mathbf{{q}}_{U},\textbf{W}^\textrm{{Lens}})\ge \sum_{k=1}^{K} {R_{j,k}^{*}}~~\forall j,
\end{align}\vspace{-0.35cm}
\end{subequations}

in which $P^{\textrm{LAS}}$ introduces the direct current (DC) power consumption of LAS-equipped transmitter (the UAV herein) before transmission of wireless signals defined in \cite{LAS2} and $P^{\textrm{Hov}}$ denotes the power consumed for UAV hovering. (\ref{QoE_max_prob.1}a) respects the power budget of the UAV ($P^{\textrm{max}}$) and (\ref{QoE_max_prob.1}b) ensures the satisfaction of minimum QoS requirement of ground users ($\Tilde{R}^{\textrm{min}}$). The motion trajectory of the UAV is lower-bounded with $\textbf{q}^{\textrm{min}}$ and upper-bounded with $\textbf{q}^{\textrm{max}}$ in (\ref{QoE_max_prob.1}c); whereas its motion speed is upper-bounded with $V^{\textrm{max}}$ in (\ref{QoE_max_prob.1}d) and finally, a common receiving data rate for each user $k$ from the $j$th beam is guaranteed in (\ref{QoE_max_prob.1}e). Stemmed from the non-convex form of (\ref{QoE_max_prob.1}b) and (\ref{QoE_max_prob.1}e) and its objective function, the optimization problem is non-convex and therefore extremely complicated to address. While convex optimization techniques %\cite{Opt1,Opt2}
impose a high complexity to solve this problem via convex transformations, the data driven learning approaches %\cite{Learn1,Learn2} 
require a large amount of sample data to this end. Instead, we propose a real-time solution with meta DRL to (\ref{QoE_max_prob.1}).
\section{Solution Strategy}\label{SS}
Inspired by the joint nature of the optimization problem (\ref{QoE_max_prob.1}), a DRL is a proper choice, which optimizes the variables jointly based on the system feedback, without requiring decomposition of the problem. However, taking the mobility of the UAV in (\ref{QoE_max_prob.1}c)-(\ref{QoE_max_prob.1}d) into account, a mismatch will occur between the training and testing stages of the DRL, if the environment change during this interval. 
\par To overcome this issue, meta learning is leveraged for initializing the training parameters of the DRL, based on various system transitions. 
First, we model the system parameters as Markov decision process (MDP). Then, a deep deterministic policy gradient (DDPG) model introduces an efficient training policy based on MDP transitions for optimizing the system variables. Eventually, meta learning is introduced to initialize the DDPG training parameters.

\subsection{Markov Decision Process}
Let us consider an edge node as the agent, responsible for capturing the system features and dynamics and also optimizing the variables. We model our wireless system in the form of MDP that reflects the environment in states, actions policies, and rewards, defined as follows.

\begin{itemize}
\item States: The state of the agent at time $t$ denoted by $\mathbf{s}(t)$, represents the interference terms in common and private data rates of RSMA for users i.e., $\mathbf{s}(t)=\left[I_{j}(t), I^{\prime}_{j}(t)\right]$. The state space $\mathcal{S}$ includes the set of all possible states.

\item Actions: The action of the agent at each time $t$ is denoted by $\mathbf{a}(t)=\left[ \mathbf{{P}}(t),\mathbf{{q}}_{U}(t),\textbf{W}^\textrm{{Lens}}(t) ,\mathbf{R}^{*}(t)\right]$ that jointly
reflects the motion trajectory, the transmit power and the lens antenna effect at the UAV and the common data rate of RSMA for users as the decision variables in (\ref{QoE_max_prob.1}), with $\mathcal{A}$ being the action space as the set of all possible actions.

\item Policy: The policy is described as the probability of each action to be taken by the agent at a given state. The policy in fact is responsible for mapping any state into a probability, corresponding to taking an action initiated from this state. So, briefly, the policy is defined as $\mathbf{\pi }_{{\Dot{\theta}}}\big( 
\mathbf{s}(t),\mathbf{a}(t)\big) =\textrm{Pr}\big( \mathbf{a}(t)|
\mathbf{s}(t)\big) $, where $\Dot{\theta}$ indicates the parameters of the agent's neural network. Accordingly, the execution process of the policy $\mathbf{\pi }_{%
{\Dot{\theta}}}$ is specified
as $\mathbf{\tau }=\left\{ \mathbf{s}(0),\mathbf{a}%
(1),...,\mathbf{s}(T-1),\mathbf{a}(T), \mathbf{s}(T)\right\} $.

\item Reward: A transition from the state $\mathbf{s}(t-1)$ to $\mathbf{s}(t)$ can be achieved by selecting the action $\mathbf{a}(t)$ at time $t$. The reward of such an action is indicated by $Re(t)=\sum\limits_{j=1}^{J}\sum\limits_{k=1}^{\left\vert
S_{j}\right\vert }R_{j,k}\left(\textbf{P}(t),\mathbf{{q}}_{U}(t),\textbf{W}^\textrm{{Lens}}(t)\right)$, which is the same objective function in (\ref{QoE_max_prob.1}).
\end{itemize}
\subsection{DDPG}
DDPG is a model-free, online, off-policy actor-critic reinforcement learning algorithm, which can be flexibly modified to work in both discrete and continuous action space, while learning a deterministic policy.
In fact, DDPG exploits a deep neural network estimator, the parameters of which should be precisely initialized before the training and adjusted while training. 
Let us denote by $%
\vartheta (s|\Delta ^{\vartheta })$ and $Q(s,a|\Delta ^{Q})$, the actor and critic networks, including corresponding parameters $\Delta ^{\vartheta }$ and $\Delta ^{Q }$, respectively. These networks are stabilized via a target actor network $\vartheta ^{^{\prime
}\vartheta ^{^{\prime }}}$ and a target critic network $Q^{\prime Q^{\prime
}}$, with corresponding parameters $\bar{\Delta}^{\vartheta }$ and $\bar{\Delta%
}^{Q}$, respectively.
Let us define a state-value
function at time $t$ as:
$Q^{\vartheta }\bigg(\bold{s}(t),\bold{a}(t)\bigg)=\mathbb{E}_{{Re}(t),\bold{s}(t+1)}\sim \mathbb{E}\left[Re\bigg(\bold{s}(t),\bold{a}(t)\bigg)+%
\tau Q^{\vartheta }\bigg( \bold{s}(t+1),\vartheta \Big( \bold{s}(t+1) \Big)\bigg) %
\right],$
in which $\tau \in \lbrack 0,1)$ stands for the discount factor.
The critic network in DDPG minimizes
$
\ell (\Delta ^{Q})=\mathbb{E}\left[ \left( Q\left(\bold{s}(t),\bold{a}(t)\Big|\Delta ^{Q}\right)
-\boldgreek{\chi}(t)\right) ^{2}\right] ,
$ as the loss function with
$
\boldgreek{\chi}(t)=Re\bigg(\bold{s}(t),\bold{a}(t)\bigg) +\tau Q\bigg(
\bold{s}(t+1),\vartheta \Big( \bold{s}(t+1)\Big) \Big|\Delta ^{Q}\bigg).
$
Similarly, the actor network aims to minimize
$
\ell(\Delta ^{\vartheta })=\mathbb{E}\Big[ Q^{\vartheta }\Big( \mathbf{s}(t),\mathbf{a}(t)\Big)
\Big|\mathbf{a}(t)=\vartheta \Big(\mathbf{s}(t)\Big|\bar{\Delta}^{\vartheta }\Big)\Big],$ as the loss function.
By opting an action $\mathbf{a}(t)\in \mathcal{A}$ at each state $\mathbf{s}(t)\in \mathcal{S}$ of the environment, the agent is transferred into the state $\textbf{s}(t+1)\in \mathcal{S}$ resulting in a reward $Re(t)$. At time $t$, the system transitions (called task from now on) are indicated by $\big[
\mathbf{s}(t),\mathbf{a}(t),\mathbf{s}(t+1),Re(t)\big]$ and stored in a replay buffer, indicated by $\mathcal{B}$. 
\subsection{Meta Learning}
%In previous section, DRL-based algorithm developed to address the joint optimization problem (\ref{}) in a UAV-enabled communications system.
By default, DRL-originated algorithms such as DDPG assume that the environment remains static during their training and testing stages. This assumption though, seems to be impractical in ever-changing environments such as UAV-enabled communication systems, regarding the non-negligible mobility of UAVs and corresponding impacts on the environment such as communication channels. Due to such a lack of adaptability, a mismatch would be possible as the testing environment follows a different distribution from the training environment. Meta learning fills this gap by improving the generalization ability to new environments and tasks. %Originally, this method was proposed in~\cite{} by incorporating the idea of the model agnostic meta learning into the DRL framework for addressing the on-policy MDP. However, regarding the off-policy MDP in our resource allocation problem, we devise an efficient meta learning model based on~\cite{} for our problem. \subsection{Definitions} The goal of this section is to train a DDPG-based DRL for quickly adapting to new tasks. To that end, 
\par \textbf{Definitions:}
The agent is trained via a variety of tasks, where each task is assumed to be an MDP and retrieved from the meta task set $\mathcal{T}=\{1,2,...,T^{\textrm{Tsk}}\}$.
%with $T^{\textrm{Tsk}}$ number of tasks. Each task is assumed to be an MDP $\big[\mathbf{s}(t),\mathbf{a}(t),\mathbf{s}(t+1),Re(t)\big]$ as stated in previous discussion. The tasks train the resource allocation policy in an optimal manner, such that a large number of states in the state space $\mathcal{S}$ and the corresponding optimized variables in the action space $\mathcal{A}$ are utilized to this purpose. 
Meta learning is of two stages, namely, meta-training and meta-adaptation. Each task $j$ in the meta-training stage has its own support set and query set, respectively expressed as $\mathbb{D}^{\textrm{Trn}}_{j}$ and $\mathbb{D}^{\textrm{Val}}_{j}$. The support set specifies the set of sampled experiences from the memory buffer ${D}_{\mathcal{T}_{j}}$ for the task $j$ and is used for updating the global parameters. The query set updates the local parameters for each task in the meta-training stage. 
\par \textbf{Meta-training Stage:} 
%The goal of this stage is to train the parameters of the actor and critic networks, such that they can adapt to a new task in an agile manner. 
In local training, each task has its own local dedicated parameters to be updated, where a set of globally initialized parameters are given. Local training is a step-by-step optimization process for each task. The global parameters, however, are updated in a periodical and synchronous fashion by contributing all local parameters and then shared among all tasks. The meta-training stage incorporates two-step updating mechanisms for updating the parameters of the actor and critic networks, devised as follows.
\subsubsection{Local Step Update} This step is done based on diverse sampled mini-batch experiences of each task from the replay memory. The optimization procedure of the local network parameters for each task $j$ can be expressed as below:
\vspace{-0.15cm}
\begin{align}\label{lossMeta}
\begin{cases}
      {{{\delta}} ^{\vartheta }_j}=\argmin_{\Delta ^{\vartheta }} \ell (\Delta ^{\vartheta},\mathbb{D}^{\textrm{Trn}}_{j}) & \forall j\in \mathcal{T} ,\\
      {{{\delta}} ^{Q }_j}=\argmin_{\Delta ^{Q }} \ell (\Delta ^{Q},\mathbb{D}^{\textrm{Trn}}_{j}) & \forall j\in \mathcal{T},
\end{cases}
\end{align}
wherein the global parameters (i.e., $\Delta ^{\vartheta}$ and $\Delta ^{\vartheta}$) are initialized and given for all tasks. Besides, ${{{\delta}} ^{\vartheta }_j}$ and ${{{\delta}} ^{Q }_j}$ introduce the local parameters of the actor and critic networks, respectively.
The loss functions in (\ref{lossMeta}) are differentiable, and therefore we can update the weights via the gradient descent method relying on the sampled experiences as the following:

\vspace{-0.35cm}
\begin{small}
\begin{align}\label{update2}
\begin{cases}
      {{{\delta}} ^{\vartheta }_{j}}^{(t)}={\delta ^{\vartheta }_{j}}^{(t-1)}-{\Tilde{\Theta}}_{\textrm{act}}\nabla_{{\delta ^{\vartheta }_{j}}^{(t-1)}} \ell ({\delta ^{\vartheta }_{j}}^{(t-1)},\mathbb{D}^{\textrm{Trn}}_{j}) & \forall j\in \mathcal{T} ,\\
      {\delta ^{Q }_{j}}^{(t)}= {\delta ^{Q }_{j}}^{(t-1)}-{\Tilde{\Theta}}_{\textrm{crt}}\nabla_{{\delta ^{Q }_{j}}^{(t-1)}} \ell ({\delta ^{Q}_{j}}^{(t-1)},\mathbb{D}^{\textrm{Trn}}_{j}) & \forall j\in \mathcal{T},
\end{cases}
\end{align}
\end{small}
\!\!with ${\Tilde{\Theta}}_{\textrm{act}}$ and ${\Tilde{\Theta}}_{\textrm{crt}}$, being the learning rate of the local step updating for actor and critic networks, respectively. More so, $t$ specifies the updating iteration, where $t=0$ corresponds to the global parameters initialized for all tasks. Once all tasks in the batch are updated, it turns to global step updating.
\subsubsection{Global Step Update}
In this step, we evaluate the adaptation ability of the updated policy for a typical task $j$, wherein the corresponding loss function is estimated over the task's query set $\mathbb{D}^{\textrm{Val}}_{j}$. Correspondingly, the overall adaptation ability of the updated policy can be achieved by aggregating the loss functions related to all tasks. The optimization procedure of the global network parameters can be stated as:
\vspace{-0.15cm}
\begin{align}\label{lossMeta2}
\begin{cases}
      {{{\Delta}} ^{\vartheta }}=\argmin_{\Delta ^{\vartheta }} \sum_{j}\ell (\delta ^{\vartheta}_j,\mathbb{D}^{\textrm{Val}}_{j}),\\
      {{{\Delta}} ^{Q }}=\argmin_{\Delta ^{Q }} \sum_{j}\ell (\delta ^{Q}_j,\mathbb{D}^{\textrm{Val}}_{j}),
\end{cases}
\end{align}
and the global updating procedure can be expressed as the following:
\vspace{-0.15cm}
\begin{align}\label{update2}
\begin{cases}
      {{{\Delta}} ^{\vartheta }}\leftarrow{{{\Delta}} ^{\vartheta }}-{{\Theta}}_{\textrm{act}}\nabla_{{{\Delta}} ^{\vartheta }} \sum_{j}\ell ({\delta ^{\vartheta }_{j}}^{(t-1)},\mathbb{D}^{\textrm{Val}}_{j}),\\
      {{{\Delta}} ^{Q }}\leftarrow {{{\Delta}} ^{Q }}-{{\Theta}}_{\textrm{crt}}\nabla_{{{{\Delta}} ^{Q }}} \sum_{j}\ell ({\delta ^{Q}_{j}}^{(t-1)},\mathbb{D}^{\textrm{Val}}_{j}),
\end{cases}
\end{align}
in which ${{\Theta}}_{\textrm{act}}$ and ${{\Theta}}_{\textrm{crt}}$ are the learning rate of the global step update. Note that by incorporating the meta learning into DRL, an additional
backward pass would be required in comparison to the conventional DRL. Eventually, once the local step and global step training are terminated, the next batch will be examined to this end.
\par \textbf{Meta-adaptation Stage:}
The trained meta model up to now, has learned how to initialize the parameters relying on generated experiences in various environments and exhibits acceptable generalization ability. In fact, the meta model can now adapt to new tasks within fewest steps. To this end, the parameters for the new step can be updated as: \vspace{-0.15cm}
\begin{align}\label{update3}
\begin{cases}
      {{{\delta}} ^{\vartheta }}={\delta ^{\vartheta }}-{\Tilde{\Theta}}_{\textrm{act}}\nabla_{{\delta ^{\vartheta }}} \ell ({\delta ^{\vartheta }}),\\
      {\delta ^{Q }}= {\delta ^{Q }}-{\Tilde{\Theta}}_{\textrm{crt}}\nabla_{{\delta ^{Q }}} \ell ({\delta ^{Q}}),
\end{cases}
\end{align}
wherein the global parameters from the meta training stage in (\ref{update2}) are used here as the initial values.
\subsection{Complexity Analysis}
The complexity of meta learning can be defined as: $\mathcal{O}\bigg(\Big(\tilde{\mathcal{B}}\mathcal{U}_{it}\mathcal{U}_{a}\mathcal{D}_{t}\mathcal{D}_{a}\Big)^{T^{\textrm{Tsk}}}\bigg),$ with $\tilde{\mathcal{B}},\mathcal{D}_{t}$ and $\mathcal{U}_{it}$, being the mini-batch size, the number of
iterations and the size of replay memory all for meta-training stage, respectively. As well, $\mathcal{U}_{a}$ and $\mathcal{D}_{a}$, respectively denote the number of iterations and the size of replay memory for meta-adaptation stage.
The complexity of DDPG is investigated from the computational and time aspects in the following.
\par \textbf{Computational Complexity:} A single-agent DDPG method employing a multi-layer perceptron (MLP), equipped with $N_{\textrm{lay}}$ number of hidden layers, has a computational complexity of
$
\mathcal{O}\bigg(\Big(N_{\textrm{epch}}\times N_{\textrm{iter}}\big(|\mathcal{S}|\times N_{\textrm{lay}}+N_{\textrm{lay}}\times |\mathcal{A}|\big)\Big)\bigg),
$
where $N_{\textrm{epch}}$ and $N_{\textrm{iter}}$ indicate the number of training epochs and the number of iterations per training epoch, respectively. 
%\par \textbf{Time Complexity:} The time complexity of DDPG can be calculated relying on floating point operations per second (FLOPS), whose order is as: $
    %\mathcal{O}\Bigg(2\sum_{i=0}^{N_{\textrm{lay}}}\bigg((2\Hat{\chi}^{\textrm{Act}}_{i}-1)\Hat{\chi}^{\textrm{Act}}_{i+1}+\Hat{\kappa}\Hat{\chi}^{\textrm{Act}}_{i+1}\bigg)+2\sum_{i=0}^{N_{\textrm{lay}}}\bigg((2\Hat{\chi}^{\textrm{Crt}}_{i}-1)\Hat{\chi}^{\textrm{Crt}}_{i+1}+\Hat{\kappa}\Hat{\chi}^{\textrm{Crt}}_{i+1}\bigg)\Bigg),$
%where at the hidden layer $i$ of the actor (critic) network, there exists a vector $\Hat{\chi}^{\textrm{Act}}_{i}$ ($\Hat{\chi}^{\textrm{Crt}}_{i}$) for dot product operation and $\Hat{\kappa}$ stands for the corresponding parameter of the activation layer.
%Thus, the overall time complexity of DDPG can be expressed as:
%$
%\mathcal{O}\bigg(\sum_{i=0}^{N_{\textrm{lay}}}\Hat{\chi}^{\textrm{Act}}_{i}\Hat{\chi}^{\textrm{Act}}_{i+1}+\sum_{i=0}^{N_{\textrm{lay}}}\Hat{\chi}^{\textrm{Crt}}_{i}\Hat{\chi}^{\textrm{Crt}}_{i+1}\bigg).
%$

	\begin{figure*}\centering
		\begin{tabular}{lccccc}
			\hspace{-0.17cm}\includegraphics[width=6.5cm,height=5cm]{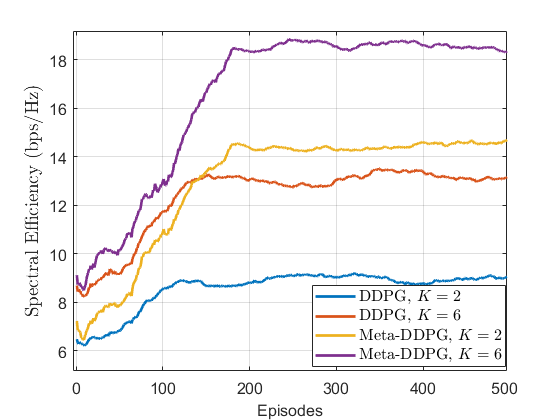}&\hspace{-0.80cm}\includegraphics[width=6.5cm,height=5cm]{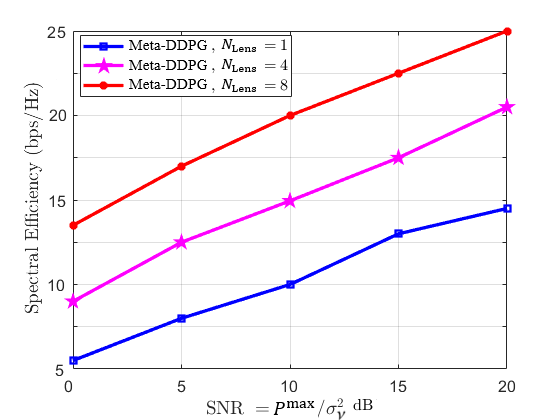}&\hspace{-0.8cm}\includegraphics[width=6.0cm,height=5cm]{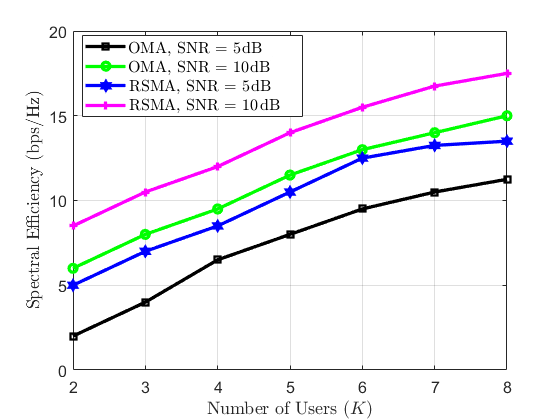}\\
			\hspace{02.050cm}(a) Convergence & \hspace{-0.6cm}(b) Number of Lenses & \hspace{-0.5cm}(c) RSMA vs. OMA\\
		\end{tabular}
		\caption{Simulation results for the proposed LAS-enabled UAV communication system.} 
		\label{fig:C}\vspace{-0.5cm}
	\end{figure*}
\section{Simulation Results}\label{SR}
In this section, we evaluate the performance of the studied LAS-empowered UAV system. The channel is assumed to be known \cite{LAS1,LAS2,LAS3} yet can be estimated separately in a future work. Therefore, the practical results would be upper-bounded by the results presented in this work. For THz band, only LoS link is taken into account, where carrier frequency is 0.2THz, the signal wavelength is $\lambda = 1.36$, the AoD angles $\theta^{l}_{k}$ and $\psi^{l}_{k}$ follow an i.i.d. uniform distribution within $[-\frac{\pi}{2},\frac{\pi}{2}]$ with $5^{\circ}$ spreading for all $l$ and $k$, while the complex-valued gain follows $\mathcal{CN}$(0,1). We set $N_{{\textrm{t}}}=32$, $N_{{\textrm{Lens}}}=4$, $N^{{\textrm{RF}}}=2$, $K=4$, $B=4$, $J=4$, $S_j=2~\forall j$, $L_k=1~\forall k$, $R^{*}_{j,k}=0.1$bps/Hz$~\forall j,k$, ${\tilde{R}^{\textrm{min}}}=1$bps/Hz, ${{P}^{\textrm{max}}}=50$Watts, $\textbf{q}^{\textrm{min}}=$(-100m,-100m), $\textbf{q}^{\textrm{max}}$=(100m,100m), ${V}^{\textrm{max}}=20$m/s and SNR=5dB, unless otherwise stated. We have modelled and initialized $P^{\textrm{LAS}}$ and $P^{\textrm{Hov}}$ from \cite{LAS3} and \cite{BU}, respectively. For the meta-DDPG method, the training parameters are $N_{\textrm{epch}}=500$, $N_{\textrm{iter}}=20$, $N_{\textrm{lay}}=4$,  $\tilde{\mathcal{B}}=64$, $\mathcal{U}_{it}=10000$, $\mathcal{U}_{a}=1000$, $\mathcal{D}_{t}=\mathcal{D}_{a}=100000$ and $T^{\textrm{Tsk}}=1000$. Moreover, 500 random channel realizations are tried for simulations.
\par Fig. \ref{fig:C}(a) illustrates the convergence of the proposed solution. The figure indicates the reward. It is observed that the proposed meta-DDPG method converges within more number of episodes than the classical DDPG method, due to massive quantities of learning
samples for better adapting to the environment. The classical DDPG method gleans faster convergence at the expense of a lower reward, because of weak generalization and difficult adaptability to our dynamic system. For instance, by considering $K=6$, we observe around 27$\%$ improvement in spectral efficiency on average, when meta learning initializes the parameters of the DDPG. Besides, the more number of users are served, the more spectral efficiency is achieved, since the proposed solution can well overcome the inter-user interference.
\par Fig. \ref{fig:C}(b) depicts how varying the values of SNR  effects the system spectral efficiency for different numbers of incorporated lenses. Observably, for better channel conditions, i.e., more SNR values, the achievable system spectral efficiency is higher. Another observation from this figure unveils that as the more number of lenses are incorporated into the LAS structure, we experience higher system spectral efficiency, thanks to better beam flexibility for steering the beam to the desired direction, especially compared to the single-lens structure\cite{BU}. 
\par From Fig. \ref{fig:C}(c), we find that the achieved spectral efficiency of our proposed scheme with RSMA, is much better than that of orthogonal multiple access (OMA)\cite{LAS3}. This is due to the non-orthogonality of RSMA, while allocating the beams to users, and also effective resource management policy for controlling the inter-user interference. For instance, by considering SNR=10dB and $K=3$, about 22\% gain is achieved over \cite{LAS3}, due to incorporating the RSMA.
\begin{table}
\centering
\captionsetup{font=small}
\captionsetup{justification=centering}
	\caption{Energy efficiency for single-lens and multi-lens UAV}
	\begin{center}
		\begin{tabular}{|l|l|l|l|l|l|}
			\hline
		\diagbox[width=14em]{EE}{SNR} & {0} & {5} & {10} & {15} & {20}\\ 
			\hline\hline
			Multi-lens UAV ($N_{\textrm{Lens}}=4$) & 1.24 & 1.32 & 1.45 & 1.61 & 1.94\\
			\hline
			Single-lens UAV ($N_{\textrm{Lens}}=1$) & 1.13 & 1.19 & 1.29 & 1.42 & 1.61\\
			\hline
		\end{tabular}
		\label{table:comp1}
	\end{center}\vspace{-0.4cm}
\end{table}
\par In Table \ref{table:comp1}, we have compared our proposed multi-lens UAV (with LAS structure) with the single-lens UAV\cite{BU}, from the energy efficiency perspective, when SNR grows. Evidently, the proposed multi-lens UAV with $N_{\textrm{Lens}}=4$ achieves a considerable energy efficiency gain over the single-lens UAV\cite{BU}. For instance, at SNR=15dB, one can observe around 13\% gain for our proposed system. What's more, the energy efficiency gap between the two baselines widens apparently, as the SNR value grows, which verifies the similar results in\cite{LAS3}.
\vspace{-0.6cm}\section{Conclusions}\label{conclutions} A green multi-antenna UAV investigated relying on LAS. We proposed a real-time resource allocation scheme based on meta reinforcement learning that achieved spectral and energy efficiency superiority compared to the single-lens UAV in literature.


\vspace{-0.3cm}\begin{thebibliography}{99}
\bibitem{3D}
L. Zhu, J. Zhang, Z. Xiao, X. Cao, D. O. Wu and X. -G. Xia, \enquote{3-D beamforming for flexible coverage in millimeter-wave UAV communications,}  \textit{IEEE Wirel. Commun. Lett.,} vol. 8, no. 3, pp. 837-840, June 2019.

\bibitem{H2}
H. Zarini, M. R. Mili, M. Rasti, S. Andreev, P. H. J. Nardelli and M. Bennis, \enquote{Intelligent analog beam selection and beamspace channel tracking in THz massive MIMO with lens antenna array,} \textit{IEEE Trans. Cogn. Commun. Netw.}, vol. 9, no. 3, pp. 629-646, June 2023.

\bibitem{H3}
H. Zarini, M. R. Mili, M. Rasti, S. Andreev and P. H. J. Nardelli, \enquote{Swish-driven GoogleNet for intelligent analog beam selection in terahertz beamspace MIMO,} in \textit{Proc. IEEE 95th Veh. Technol. Conf. (VTC2022-Spring)}, Helsinki, Finland, 2022, pp. 1-6.

\bibitem{H4}
H. Zarini, M. R. Mili, M. Rasti, P. H. J. Nardelli and M. Bennis, \enquote{Xavier-enabled extreme reservoir machine for millimeter-wave beamspace channel tracking,} in \textit{Proc. IEEE Wirel. Commun. Netw. Conf. (WCNC)}, Austin, TX, USA, 2022, pp. 1683-1688.

%\bibitem{M1}
%D. Xu, Y. Sun, D. W. K. Ng, and R. Schober, \enquote{Multi-user MISO UAV communications in uncertain environments with no-fly zones: Robust trajectory and resource allocation design,} \textit{IEEE Trans. Commun.,} vol. 68, no. 5, pp. 3153–3172, May 2020.


%\bibitem{M3}
%S. K. Singh, K. Agrawal, K. Singh, A. Bansal, C. -P. Li and Z. Ding, \enquote{On the Performance of Laser-Powered UAV-Assisted SWIPT Enabled Multiuser Communication Network With Hybrid NOMA,} \textit{IEEE Trans. Commun.,} vol. 70, no. 6, pp. 3912-3929, June 2022.

\bibitem{BU}
Z. Chen, N. Zhao, D. K. C. So, J. Tang, X. Y. Zhang and K. -K. Wong, \enquote{Joint altitude and hybrid beamspace precoding optimization for UAV-enabled multiuser mmWave MIMO system,} \textit{IEEE Trans. Veh. Technol.}, vol. 71, no. 2, pp. 1713-1725, Feb. 2022.

%\bibitem{D1} T. Xie, L. Dai, D. W. K. Ng, and C.-B. Chae, \enquote{On the power leakage problem in millimeter-wave massive MIMO with lens antenna arrays,} \enquote{IEEE Trans. Signal Process.,} vol. 67, no. 18, pp. 4730–4744, Sep. 2019.

%\bibitem{D2} H. Tataria, M. Matthaiou, P. J. Smith, G. C. Alexandropoulos, and V. F. Fusco, \enquote{Impact of RF processing and switching errors in lens-based massive MIMO systems,} in \textit{Proc. IEEE 19th Int. Workshop Signal Process. Adv. Wireless Commun. (SPAWC),} Jun. 2018, pp. 1–5.

\bibitem{LAS1} M. Karabacak, H. Arslan, and G. Mumcu, \enquote{Lens antenna subarrays in mmWave hybrid MIMO systems,} \textit{IEEE Access}, vol. 8, pp. 216634–216644, 2020.

\bibitem{LAS2}
L. Afeef, G. Mumcu and H. Arslan, \enquote{Energy and spectral-efficient lens antenna subarray design in MmWave MIMO systems,} \textit{IEEE Access}, vol. 10, pp. 75176-75185, 2022.

\bibitem{LAS3}
S. Cetinkaya, L. Afeef, G. Mumcu and H. Arslan, \enquote{Heuristic inspired precoding for millimeter-wave MIMO systems with lens antenna subarrays,} in \textit{Proc. IEEE 95th Veh. Technol. Conf. (VTC2022-Spring),} Helsinki, Finland, 2022, pp. 1-6.

\bibitem{RSMA}
M. Mert Sahin, O. Dizdar, B. Clerckx, H. Arslan, \enquote{Multicarrier rate-splitting multiple access: superiority of OFDM-RSMA over OFDMA and OFDM-NOMA,} 2023, \textit{arXiv:2303.14540v1}.

\bibitem{MIMO1}
Z. Wang, T. Lv, J. Zeng and W. Ni, \enquote{Placement and resource allocation of wireless-powered  multiantenna UAV for energy-efficient multiuser NOMA,} \textit{IEEE Trans. Wirel. Commun.}, vol. 21, no. 10, pp. 8757-8771, Oct. 2022.

\bibitem{MIMO2}
W. Mao, K. Xiong, Y. Lu, P. Fan and Z. Ding, \enquote{Energy consumption minimization in secure multi-antenna UAV-assisted MEC networks with channel uncertainty,} \textit{IEEE Trans. Wirel. Commun.}, vol. 22, no. 11, pp. 7185-7200, Nov. 2023.

\bibitem{MIMO3}
M. T. Mamaghani and Y. Hong, \enquote{Intelligent trajectory design for secure full- duplex MIMO-UAV relaying against active eavesdroppers: A model-free reinforcement learning approach,} \textit{IEEE Access,} vol. 9, pp. 4447-4465, 2021.

\bibitem{meta}
Y. Eghbali, \textit{et. al.,}, \enquote{Beamforming for STAR-RIS-Aided Integrated Sensing and Communication Using Meta DRL,} \textit{IEEE Wirel. Commun. Lett.}, vol. 13, no. 4, pp. 919-923, Apr. 2024.
%\bibitem{LAS3} M. Karabacak, G. Mumcu, and H. Arslan, \enquote{Hybrid MIMO architecture using lens arrays,} U.S. Patent 10 714 836, Jul. 14, 2020.

%\bibitem{LAS2} G. Mumcu, M. Kacar, and J. Mendoza, \enquote{Mm-wave beam steering antenna with reduced hardware complexity using lens antenna subarrays,} \textit{IEEE Antennas Wireless Propag. Lett.,} vol. 17, no. 9, pp. 1603–1607, Sep. 2018.

%\bibitem{Zhang}
%Y. Zeng and R. Zhang, \enquote{Millimeter Wave MIMO With Lens Antenna Array: A New Path Division Multiplexing Paradigm,} \textit{IEEE Trans. Commun.}, vol. 64, no. 4, pp. 1557-1571, April 2016.

%\bibitem{Opt1}
%H. Zarini, N. Gholipoor, M. R. Mili, M. Rasti, H. Tabassum and E. Hossain, \enquote{Resource Management for Multiplexing eMBB and URLLC Services Over RIS-Aided THz Communication,} \textit{IEEE Trans. Commun.}, vol. 71, no. 2, pp. 1207-1225, Feb. 2023.


%\bibitem{Opt2}
%H. Zarini, A. Khalili, H. Tabassum and M. Rasti, \enquote{Joint Transmission in QoE-Driven Backhaul-Aware MC-NOMA Cognitive Radio Network,} \textit{IEEE Global Commun. Conf.:(GLOBECOM)}, Taipei, Taiwan, 2020, pp. 1-6.

%\bibitem{Learn1}
%H. Zarini, N. Gholipoor, M. R. Mili, M. Rasti, H. Tabassum and E. Hossain, \enquote{Liquid State Machine-Empowered Reflection Tracking in RIS-Aided THz Communications,} \textit{IEEE Global Commun. Conf.:(GLOBECOM)}, Rio de Janeiro, Brazil, 2022, pp. 5273-5278.

%\bibitem{Learn2}
%H. Zarini, A. Khalili, H. Tabassum, M. Rasti and W. Saad, \enquote{AlexNet Classifier and Support Vector Regressor for Scheduling and Power Control in Multimedia Heterogeneous Networks,} \textit{IEEE Trans. Mobile Comput.}, vol. 22, no. 5, pp. 2520-2536, 1 May 2023.
\end{thebibliography}
\end{document}